\documentclass[a4paper,aps,twocolumn,floats,nofootinbib]{revtex4-1}
\usepackage{graphics,graphicx,epsfig, wrapfig}
\usepackage{amssymb,color}
\pdfoutput=1
\usepackage{epsf,epstopdf,wrapfig, layouts}
\usepackage {amsmath}

\usepackage{sidecap}

\usepackage{jabbrv} 
\newcommand{\beq}{\begin{equation}}
\newcommand{\eeq}{\end{equation}}
\newcommand{\beqn}{\begin{eqnarray}}
\newcommand{\eeqn}{\end{eqnarray}}

\definecolor{M}{rgb}{1,0.5,0}	% color for comments from M

\begin{document}

\title{Collective behavior of place and non--place neurons in the hippocampal network}

%\preprint{APS/123-QED}
%
%\title{Pairwise interactions capture collective information in hippocampal place and silent cells}% Force line breaks with \\

\author{Leenoy Meshulam,$^{1,2,3}$ Jeffrey L. Gauthier,$^{1}$ Carlos D. Brody,$^{1,4,5}$ David W. Tank,$^{1,2,4}$ and William Bialek$^{2,3}$}

\affiliation{$^1$Princeton Neuroscience Institute, $^2$Joseph Henry Laboratories of Physics, $^3$Lewis--Sigler Institute for Integrative Genomics,  $^4$Department of Molecular Biology,  and $^5$Howard Hughes Medical Institute, Princeton University, Princeton, NJ 08544}

%\date{\today} 

\begin{abstract}

	\noindent Discussions of the hippocampus often focus on place cells, but many neurons are not place cells in any given environment. Here we describe the collective activity in such mixed populations, treating place and non--place cells on the same footing.  We start with optical imaging experiments on CA1 in mice as they run along a virtual linear track, and use maximum entropy methods to approximate the distribution of patterns of activity in the population, matching the correlations between pairs of cells but otherwise assuming as little structure as possible.  We find that these simple models accurately predict the activity of each neuron from the state of all the other neurons in the network, regardless of how well that neuron codes for position.  These and other results suggest that place cells are not a distinct sub--network, but part of a larger system that encodes, collectively, more than just place information.  
	
\end{abstract}

\maketitle

\section{Introduction}

The hippocampal formation plays a key role in various memory processes, and is crucial for spatial navigation.  In humans, individual hippocampal neurons whose activity is associated with specific items or episodes, were found sparsely among a larger population of non-responsive neurons \cite{QuirogaReddyKreimanEtAl2005,waydo2006sparse,Gelbard-SagivMukamelHarelEtAl2008}. In rodents, cells coding for place, velocity, and head direction have been recorded from the hippocampus and its neighboring structures \cite{McNaughtonBarnesOkeefe1983,SargoliniFyhnHaftingEtAl2006, HaftingFyhnMoldenEtAl2005}. One well identified group of hippocampal neurons are \textit{place cells}  \cite{OKeefe1971}.  These cells are active selectively whenever the animal visits a particular location, i.e. each of these neurons has a specific \textit{place field} in the environment. Typically, once the animal is familiar with a given environment, about 30\% of CA1 pyramidal neurons will fire with significant place fields, and the activity of these cells constitutes a cognitive map  \cite{OKeefe1971} that spans the full environment. Meanwhile, the remaining cells, often referred to as \textit{silent cells} \cite{Thompson1989}, fire with little to no spatial selectivity, and usually are less active over all \cite{OkeefeConway1978, OKeefeNadel1979,Thompson1989,KubieMuller1991,LeutgebLeutgebTrevesEtAl2004}. While some neurons qualify inarguably as place cells, with strong, reliable place fields, for others the boundary between ``place'' and ``non--place'' is arbitrary. 

\begin{figure*} 
	
	\includegraphics {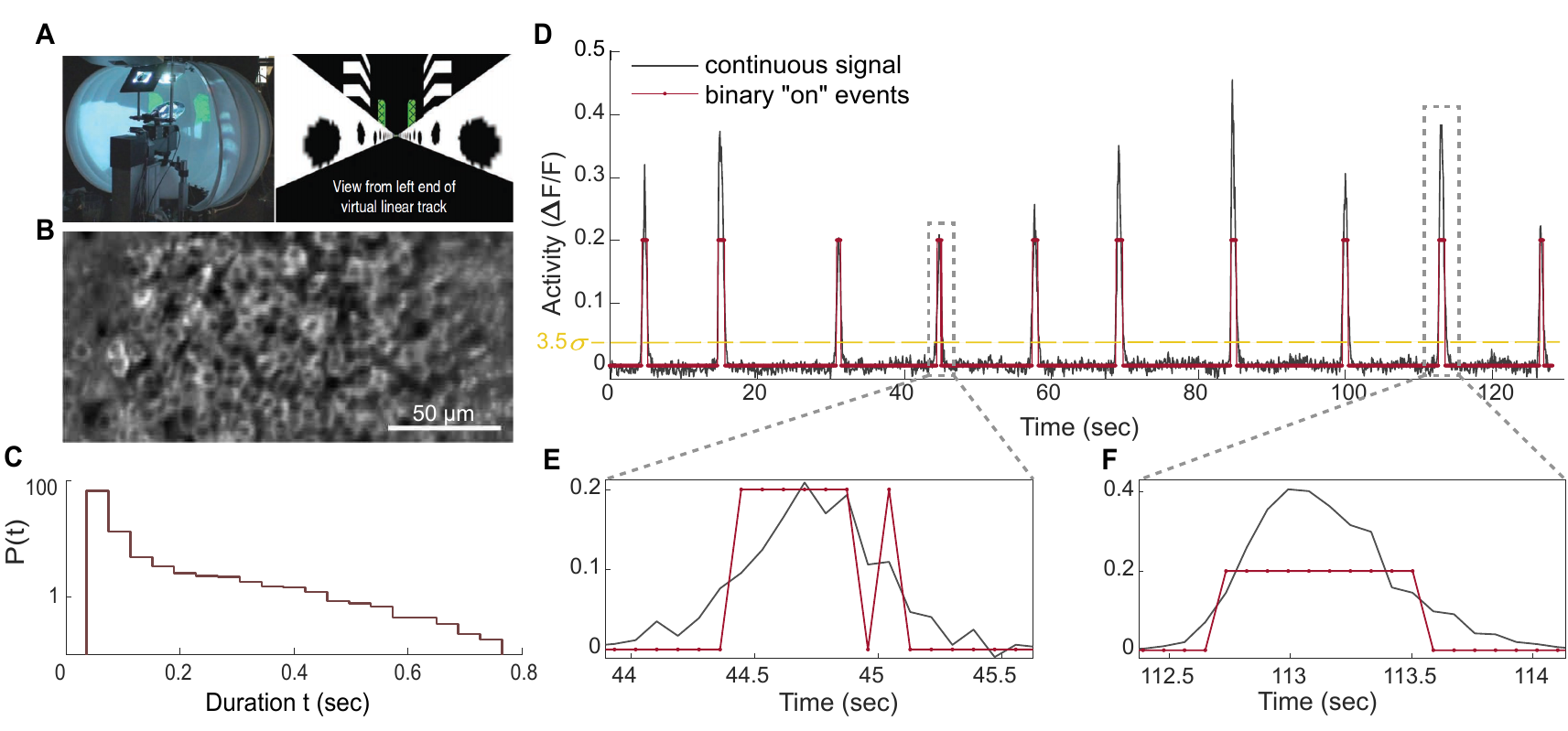}
	
	\caption{\textbf{Experimental setup} \textbf{(A)} Left panel shows a photograph of the experimental apparatus, consisting of spherical treadmill, a virtual reality apparatus, and a custom two-photon microscope. Right panel shows a top view of the virtual environment of a linear track. (Adapted from \cite{Dombeck2010}). \textbf{(B)} Field of view under the microscope. Scale bar is 50$\mu$m. \textbf{(C)} Identifying significant calcium transients: in gray, raw signal from a neuron in units of normalized fluorescence. In red (arbitrary height), the binarization of the signal: the neuron is ``on'' (which is converted to a ``1'' value) in every time point the red line is not 0. Neuron is ``off'' (which is converted to a ``0'' value) in every time point the red line is 0 (see Appendix \ref{cont_bin} for more details). Dashed yellow horizontal line shows the 3.5$\sigma$ threshold in the distribution of all fluorescence values. Points above this threshold qualify in our initial step of binarization of the signal as potential activity events. \textbf{(D)} Distribution of the ``on'' activity events' durations, on a log scale. \textbf{(E+F)} Zoom in examples of two raw activity transients (gray) and their binarized version (red). \label{data1}}
\end{figure*}

One way of interpreting the phenomenology of place cells is that individual cells respond independently to the location of the animal at any moment in time. This approach stems from the view of pyramidal cells responding to specific combinations of complex features of the sensory environment. However, activity in CA1 seems to be more complex than a simple evoked response to external stimulus, and cannot all be accounted for by this ``sensory features'' view. A different approach is to treat the cells as an interacting network, asking how their collective behavior can give rise to place modulated activity. The latter idea has been widely employed in theoretical models of the hippocampus, which suggest attractor dynamics as the potential mechanism underlying three of the main functions of hippocampus: memory storage, pattern separation, and pattern completion \cite{SamsonovichMcNaughton1997, RedishTouretzky1999, KaliDayan2000, FuhsTouretzky2006}. 

A large fraction of hippocampal studies focus on sub-regions CA1 and CA3, and in particular on the sequential activity exhibited by their neurons’ spatial (and non-spatial) fields.  While it has been suggested that CA1 contributes to the order of a spatial memory sequence, it is not yet known if CA1 is involved in the integration of sequential elements of a memory, or in bridging events that happen sequentially during an episode. \cite{kesner2005role,hunsaker2008evaluating,farovik2010distinct}. Although many of the models for hippocampal circuitry refer to CA3 as the main attractor network in hippocampus \cite{treves1992computational,jensen1996hippocampal,levy1996sequence,battaglia1998attractor,lisman1999relating,ColginLeutgebJezekEtAl2010}, other computational studies suggest that even though CA1 has less recurrent connections than CA3, it might still function as a continuous attractor, maintainning the continuity of sequential activity \cite{DroulezBerthoz1991,TsodyksSejnowski1995}. 

While some experimental studies have tried to characterize the instability in the place code and to probe the activity of the cells whose activity is not significantly place modulated \cite{Fenton1998,EpszteinBrechtLee2011, FergusonJacksonRedish2011,kelemen2013organization}, little to no theoretical effort has been dedicated to study the phenomenology of CA1 beyond position encoding. Furthermore, recent development of large-scale neural imaging technologies now allows us access to the activity of all neurons -- place and non--place coding -- in a field of view. With this progress arises the need for modeling approaches that will help elucidate these large population--level codes. 

Among the possible mathematical models that can reproduce a given pattern of neural activity, we chose the one that does so without incorporating additional structure or assumptions. This minimally informative model is the one that maximizes the entropy of the system \cite{Jaynes1957,Jaynes1982}.  In this study, we use the maximum entropy approach to build a model for the probability of all different joint activity patterns of the neurons. We report a successful description of the full probability distribution inferred solely from mean activities and pairwise correlations, which yields accurate predictions both for higher-order phenomena in the network and for the activity of individual place and non-place neurons in relation to the rest of the network. Finally, we show that this collective description for the population as a whole yields better predictions than the classical view of individual place cells as independently encoding locations on a spatial map.

\section{Results}
\section*{Cellular resolution imaging of neural activity in the mouse during runs along a virtual track}
\label{sec_data}

We analyze data taken from transgenic mice that express a genetically encoded calcium indicator GCaMP3. Our virtual reality setup allows imaging neural activity with cellular resolution in awake, head-restrained mice while they run on a spherical treadmill (Fig 1A). Each imaging session included up to ~80 simultaneously active neurons in CA1 hippocampus, recorded as a mouse runs along a 4-meter-long virtual linear track. Right before the end of the track the mouse received a water reward. Runs were consecutive in nature because once the mouse reached the end of the virtual reality environment, the next one started immediately. In agreement with the literature, sub-populations of the imaged neurons (usually $\sim30$) were found to be place cells, whose activity occur in specific place fields along the environment. (Fig \ref{data2}C, and Fig \ref{examps} in Appendix \ref{def_place_cells}).

The raw data from this experiment is essentially a movie. To reduce this to activity of individual neurons as a function of time, we follow the pre-processing steps described in the Experimental Procedures. In outline, we correct the movies for motion of the brain, identify each neuron as a “region of interest” in the movie, verify these regions manually, and then associate the activity of each cell with the integral of the fluorescence signal over the corresponding region. We normalize the fluorescence traces extracted from each neuron, and set a threshold above which all maxima identified are recognized as potential activity events; this threshold is set to $\sim 3.5\sigma$, where $\sigma$  is the standard deviation of the distribution of all fluorescence values of that session (dashed horizontal yellow line in Fig 1D). Next, we discard transients whose shape contradicts a rising-decaying pattern typical of neural activity (Appendix \ref{cont_bin}. Finally, we binarize the time series: a neuron state $\sigma_{\rm i}$ is defined as 1 for every time bin during an identified activity event, and as 0 for the rest (and most) of the time, when it is silent, as shown in Fig 1C. Since each point in our field of view is scanned every 70 ms, we use the same discretization as our effective ``time bin'' to binarize the signal. Panels E and F of Fig 1 depict two events and their binarized versions. A closer look at panel E reveals a case where, while the event was clearly identified as significant activity, our algorithm found a part of the decay slope which best fits a burst of activity followed by a single event rather than one unified burst of activity, since the fluorescence level falls too quickly to be explained by continuous firing. For more details about the final assignment of 0 and 1 values see \ref{fit} in Appendix \ref{app_learning}.
Analysis was performed on all concatenated frames from single sessions, which included any time period the mouse was running, as well as the inter-trial intervals and any time points where the mouse was not engaged in the task if such existed (rare). Data were used as a single stream, and no behavior-related information was incorporated into the model other than for validation or visualization purposes.

\begin{figure} 
	\includegraphics {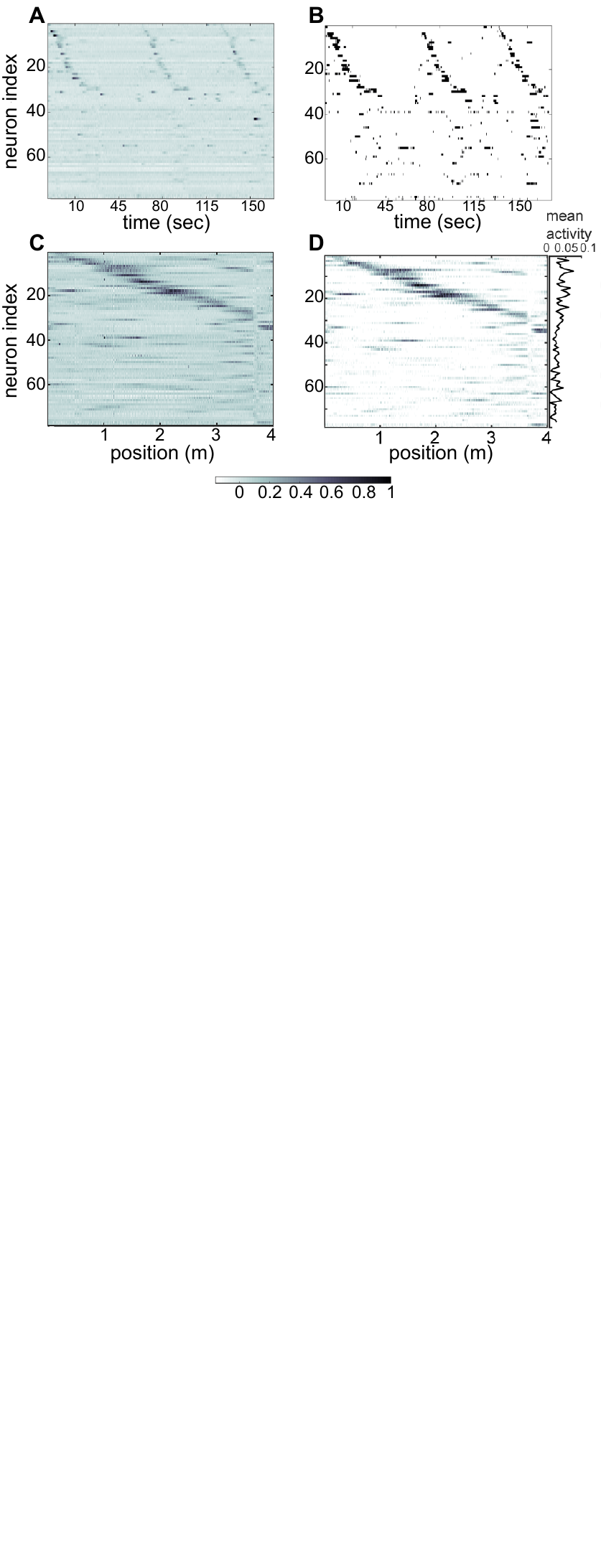}
	   \vspace{-430pt}
	\caption{\textbf{Place and non--place related activity }  \textbf{(A)} Three consecutive runs down the linear track. During each traversal of the environment, 32 out of the 78 cells imaged in the field of view exhibit place modulated activity. Place cells were sorted based the averaged activity's center of mass (panel C). This is the thresholded continuous fluorescence signal as in Fig \ref{data1}C.  \textbf{(B)} Binarized version of panel A. The sequential nature of the place activity is preserved during the discretization process. \textbf{(C)} Continuous neural activity averaged across runs, calibrated against position. Sorted based on center of mass.  \textbf{(D)} Binarized neural activity averaged across runs, calibrated against position. On the right, mean activity values corresponding to the neurons on the left. Non-place cells are usually less active than place cells, yet rarely silent. \label{data2}}
\end{figure}

\section*{The maximum entropy model}
\label{sec_maxent}

A network of N neurons has $\Omega=2^N$ possible states, even if we limit ourselves to describing each cell as being active or silent, $\sigma_{\rm i} = 1$ or $0$, respectively, for each neuron $\rm i$.  With $N=78$ cells in our experiment, this number of possible states is  $\Omega \sim 3\times 10^{23}$, a genuinely astronomical number.  We expect that neural activity is organized, and so the network does not wander at random among all these possibilities. Although there are many ways that we could try to characterize this organization, perhaps the simplest approach is to ask for the probability distribution over all the possible states, $P(\sigma_{\rm i})$: how likely is it that we will find any particular combination of active and inactive neurons at any one moment in time?
A probability distribution over $\Omega$ states is just a list of $\Omega$ numbers that sum to one. In large networks, it is not possible to conceive of an experiment that would allow us to measure all these numbers. Thus, any attempt to make a ``probabilistic model'' of neural activity must rest on some (dramatically) simplifying hypotheses.  As emphasized long ago by Jaynes \cite{Jaynes1957,Jaynes1982}, the maximum entropy construction provides a path for building models in which simplifying hypotheses are explicit and testable.
  
The idea of the maximum entropy method is to build models that match certain experimental facts exactly, but otherwise have as little structure as possible. Thus, if we draw states out of the probability distribution, these combinations of activity and silence across this network will be as random as they can while still matching, on average, the experimental facts that we have chosen as being crucial. The choice of what to match embodies our intuition, and the shorter the list of facts the more drastic our simplification.  We emphasize that there is no obviously correct choice for which facts to match: each choice represents a different simplifying hypothesis, and must be tested, quantitatively. 
 
In trying to choose which facts to match, we are looking for a relatively small set of things we can measure, reliably, which are sufficient to capture the way in which the patterns of activity in the network are ordered.  As an example, we can get a very accurate model for the joint distribution of velocities of all the birds in a flock by matching the correlations between individual birds and their near neighbors; the global ordering of the flock, the correlations among fluctuations in the velocities of distant birds, and even higher-order correlations among multiple birds, all emerge from propagation of the near-neighbor correlations, as predicted by the maximum entropy model \cite{Bialek2012,Bialek2014}.
Previous work on maximum entropy approaches to the description of neural activity has explored different possibilities for what should be matched.  At one extreme, the network might be characterized by the fact that certain specific combinations of activity and silence occur with surprisingly high frequency, and we could match these frequencies for the most common combinations \cite{Ganmor2011a}.  At the opposite extreme, we could focus on measures of global activity, and match the frequency with which we see $K$ out of $N$ neurons being active simultaneously, without regard to their identity \cite{tkavcik2013simplest}.  Here we pursue the original idea of matching the mean activity of individual neurons, and the correlations between pairs of neurons \cite{Schneidman2006,Tkacik2006,tkacik2009spin}.

We expect that networks with different mean levels of activity will behave differently, and so we insist that any model of the network as a whole match the average activity of each neuron. For each neuron $\rm i$, we have $\sigma_{\rm i} = 1$  when that neuron is active, and  $\sigma_{\rm i} = 0$  when it is silent; matching the mean activity of each cell is the statement that if we compute the mean in our model probability distribution $P(\sigma_{\rm i})$ we get the same answer as from experiment: 

\begin{equation}
\sum_{\{\sigma_{\rm i}\}} P(\{\sigma_{\rm i}\}) \sigma_{\rm j} = \langle \sigma_{\rm j}\rangle_{\rm expt} ,
\label{match_mean}
\end{equation}
where $\langle \cdots \rangle_{\rm expt}$ is the average computed from the experimental data. Similarly, we know that correlations between pairs of neurons capture some aspects of the network's coherent activity, and so we will test the hypothesis that matching these correlations is sufficient to predict all the possible higher--order structures.  Mathematically, the pairwise correlation between neuron $\rm i$ and neuron $\rm j$ can be defined either as a covariance,  

\begin{equation}
C_{\rm ij} = \langle \left( \sigma_{\rm i} - \langle \sigma_{\rm i}\rangle \right)
\left( \sigma_{\rm j} - \langle \sigma_{\rm j}\rangle \right)\rangle ,
\label{Cij_def}
\end{equation}
or as a correlation coefficient,
\begin{equation}
c_{\rm ij} = {{C_{\rm ij}}\over
{\sqrt{C_{\rm ii} C_{\rm jj}}}} .  
\label{cij_def}
\end{equation}
To match this entire matrix of correlations, we need to be sure that our model matches the matrix of second moments,
\begin{equation}
\sum_{\{\sigma_{\rm i}\}} P(\{\sigma_{\rm i}\}) \sigma_{\rm j} \sigma_{\rm k}= \langle \sigma_{\rm j}\sigma_{\rm k}\rangle_{\rm expt} .
\label{match_pair}
\end{equation}

\begin{figure} %{r}{1\textwidth}
	
	\includegraphics{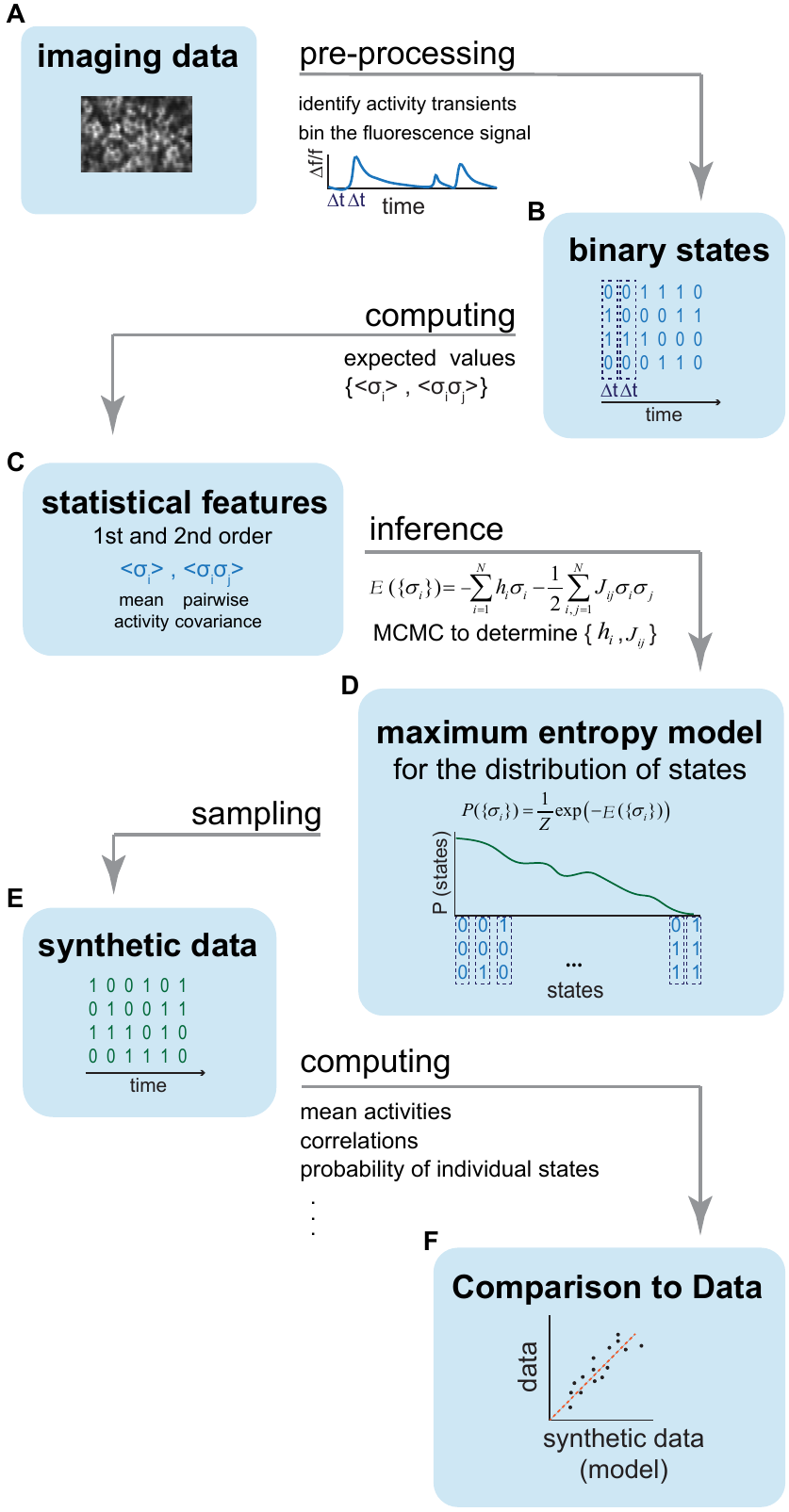}
	
	\caption{\textbf{Steps in building the maximum entropy model}  \textbf{(A)} A typical field of view in CA1 hippocampus. Initial steps include motion correction and identification of individual cells. \textbf{(B)} After discretizing the signal from each neuron, we now have a matrix of all concatenated trials for all neurons were each neuron was assigned a ``0'' or ``1'' value for every moment in time. \textbf{(C)} Compute the statistical features of the data to which we are going to fit the model. We require the model to match exactly the first and second moments. \textbf{(D)} Finding the values of a set of $h_{\rm i}$ and $J_{\rm ij}$ that match the statistical features of the data via Markov Chain Monte Carlo simulation of the model. The result is a full probability distribution, such that every possible population state is assigned a specific probability to occur. \textbf{(E)} Sample population states from the inferred distribution to obtain a matrix of synthetic data. \textbf{(F)} To test whether the inferred distribution is a good model, compare the same measures we computed on the real data to those computed on the synthetic data. \label{steps}}
	
\end{figure} 

There are infinitely many probability distributions that are consistent with the constraints in Eqs (\ref{match_mean}) and (\ref{match_pair}).  Among these, we want to choose the model that has the least possible structure, or equivalently generates the most random possible states.   Although such a characterization may seem vague, Shannon proved that the only consistent way to measure the (lack of) structure in a probability distribution, or the degree of randomness, is to compute the entropy of the distribution,
\begin{equation}
S = - \sum_{\{\sigma_{\rm i}\}} P(\{\sigma_{\rm i}\}) \log  \left[ P(\{\sigma_{\rm i}\}) \right].
\label{entropy_def}
\end{equation}
Concretely, then, we want to find the probability distribution that maximizes $S$ while obeying the constraints from Eqs (\ref{match_mean}) and (\ref{match_pair}).  The formal solution to this constrained optimization problem is
\begin{eqnarray}
 P(\{\sigma_{\rm i}\}) &=& \frac{1}{Z}\exp[- E(\{\sigma_i\})]
  \label{model1}\\
E(\{\sigma_i\}) &=& -\sum_{{\rm i} =1}^{N} h_{\rm i}\sigma_{\rm i} -\frac{1}{2}\sum_{{\rm i},{\rm j}=1}^{N}J_{\rm ij}\sigma_{\rm i}\sigma_{\rm j }.  
\label{model2}
 \end{eqnarray}
We note that this is mathematically equivalent to the Boltzmann distribution for a set of Ising spins $\sigma_{\rm i}$ that experience external magnetic fields $h_{\rm i}$ and interact with one another through couplings $J_{\rm ij}$ \cite{mezard1990spin}.  The partition function $Z$ serves to normalize the distribution,
\begin{equation}
Z = \sum_{\{\sigma_{\rm i}\}} \exp[- E(\{\sigma_i\})].
\end{equation}

This formal solution for the probability distribution involves parameters $\{h_{\rm i}, J_{\rm ij}\}$, and there are exactly as many of these parameters as we have constraints from matching the mean activity and pairwise correlations that we see in the population;   in principle, then, parameters of our model are determined exactly by solving Eqs (\ref{match_mean}) and (\ref{match_pair}).  In practice, we can't find the exact expectation values in our model, and so we compute these (as is standard in statistical physics) by Monte Carlo simulation.  Thus we ``solve'' Eqs (\ref{match_mean}) and (\ref{match_pair}) by using Monte Carlo to generate an estimate of the left hand side of the equation, comparing to the right hand side (from the data), and then adjusting the parameters until we get closer to equality; for details see Appendix \ref{app_learning}. Our strategy, going from raw fluorescence data to the model, is summarized in Fig \ref{steps}.

\section*{Maximum entropy model for a population of hippocampal neurons} 

\begin{figure} %{r}{1\textwidth}
	
\includegraphics {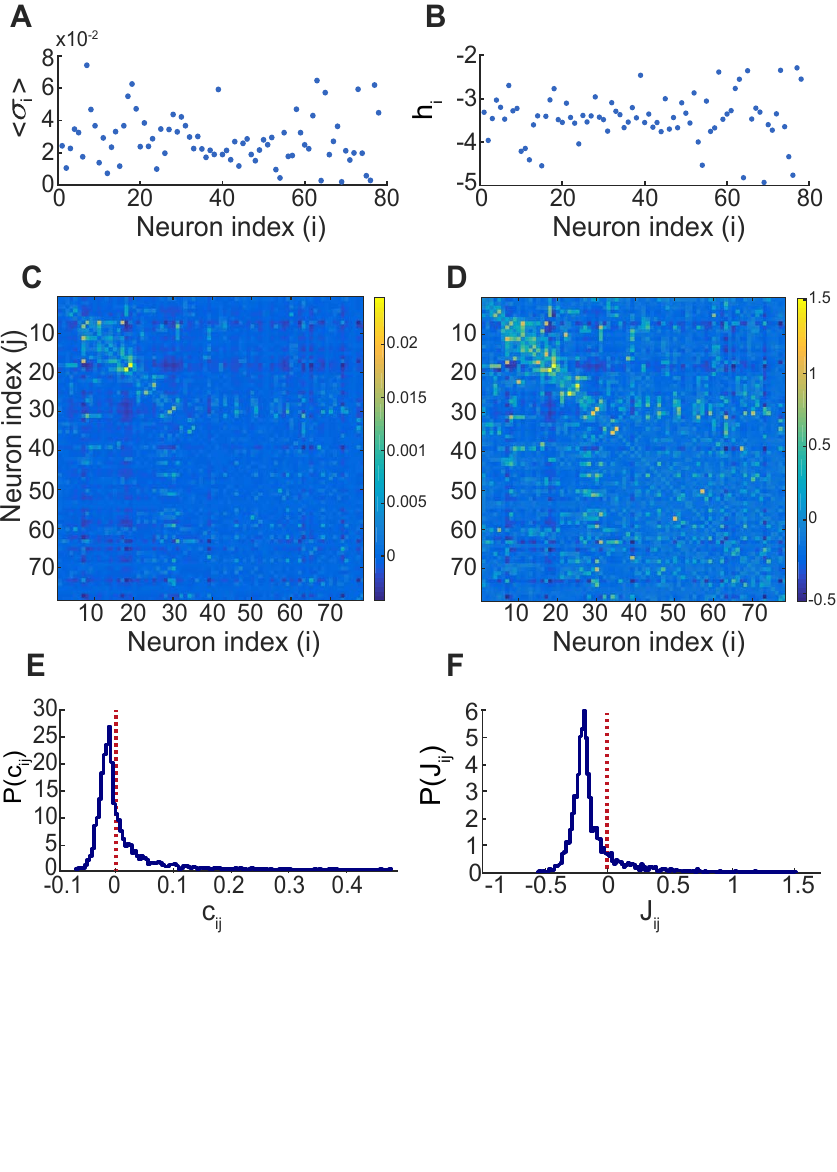}
\vspace{-65pt}
\caption{\textbf{Learning the maximum entropy model.}  \textbf{(A)} Mean activity, $\langle \sigma_i \rangle_{\rm expt}$, for each one of the neurons, computed from the data. \textbf{(B)} Coefficients (magnetic fields) $h_{\rm i}$, in the model; positive values bias the cell to be active. \textbf{(C)} Probability distribution of the correlation coefficients in the data, $c_{\rm ij}$ for $\rm i \neq \rm j$ as defined in Eq. \ref{Cij_def}. Note the peak of slightly negative values indicating a significant population of neurons being weakly negatively correlated, as expected from place cells, whose firing is mostly orthogonal to each other. \textbf{(D)} Probability distribution of the coefficients (coupling constants), $J_{\rm ij}$, obtained after fitting the model. \textbf{(E)} Pairwise covariances, $C_{\rm ij}$, as defined in Eq. \ref{cij_def}, computed from the data. $C_{\rm ij}$ was set to 0 for ease of visualization. (F) Coupling constants of the model, $J_{\rm ij}$. Positive couplings favor positively correlated activity. $J_{\rm ii}$ is redundant with $h_{\rm i}$ and is set to 0. \label{basics}}
\end{figure}

We apply the maximum entropy approach to the experimental data on 78 CA1 neurons (\ref{data2}).  The crucial experimental quantities are the mean activities of each neuron, shown in Fig \ref{basics}A, and the pairwise correlations, illustrated in Figs \ref{basics}C and E.  Because we have described activity by a binary variable, the man activity measures the fraction of time bins in which a neuron is ``on'' and we see that this is quite small ($\sim 0.03$), as expected. Corresponding to these low probabilities, the parameters $h_{\rm i}$ in our model [Eq (\ref{model2})] are relatively large and negative.

As noted in Fig \ref{data2}, a significant fraction of cells in the population are place cells.  For these, we expect positive correlations between cells with overlapping place fields, and negative correlations between cells with non-overlapping place fields.  More generally we see that the distribution of correlation coefficients peaks, and has the bulk of its weight, at $c_{\rm ij}$ small and negative (Fig \ref{basics}E), and the covariance matrix of the activity is dominated by small negative terms far from the diagonal (Fig \ref{basics}C).  Corresponding to these experimental results, the coupling constants $J_{\rm ij}$ in the model also have a distribution peaking at small negative values (Fig \ref{basics}F), and if we view these coupling as a matrix we see widespread, weakly negative terms away from the diagonal, with a small set of positive terms clustered near the diagonal (Fig \ref{basics}D).

The basic properties that we see in these data are reproduced in five other data sets, from different mice running through the same virtual track (Appendix \ref{more_expts}). There are uncertainties in the parameters, but no sign that we are overfitting, since learning the model from a limited fraction of the data predicts mean activities and correlations that agree with those computed from the remainder of the data (Fig \ref{fit} in Appendix \ref{app_learning}).

\section*{Model predictions}

Maximum entropy models are, by definition, the least structured models that can match particular experimental facts. But the set of facts we choose to match---here, the mean activity and pairwise correlations of the neurons---is small compared to the space of possible states for the network, and there is no guarantee that these few measured quantities provide a sufficient characterization of the system.  What is crucial, however, is that everything else that we can predict about the behavior of the network from the model involves no new parameters.  Thus, what follows are predictions, not fits.

\begin{figure}

    \includegraphics{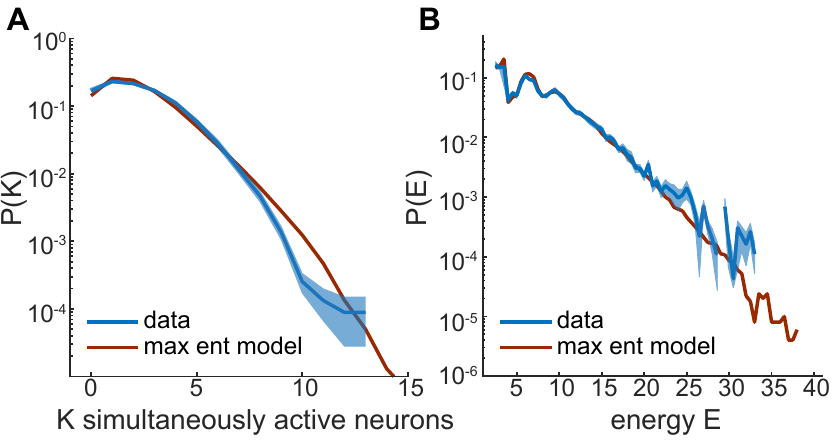}

  \caption{\textbf{Model predictions.} \textbf{(A)} The probability that $K$ out of the $N=78$ neurons in the population are active simultaneously. Data (blue) and predictions from the model (red). \textbf{(B)} The distribution of effective energies, or log probabilities, that the model assigns to every possible state in the network.  In blue, the distribution over states as they occur in the experiment.  In red, the distribution predicted from the model itself.  Data in both panels shown as mean and standard deviations over random halves of the data. \label{test1}}
\end{figure} 

Although the mean probability of a single cell being active is small (Fig \ref{basics}), there are many neurons in the population, and so it is possible to observe not just a single cell being active but many cells being active simultaneously.  While we do not have enough data to estimate the probability of every distinct ``word'' built out of combinations of activity and silence, we can estimate the probability $P_N(K)$ that $K$ out of the $N$ cells in the network are active in the same small window of time.  As shown in Fig \ref{test1}A, there is significant probability out to $K\sim 8$, and we can follow the small probabilities out to $K=12$.  Across most of this range, the model makes correct predictions, within the (small) error bars of the measurement.  

\begin{figure*}  
\centerline{\includegraphics[width=\linewidth] {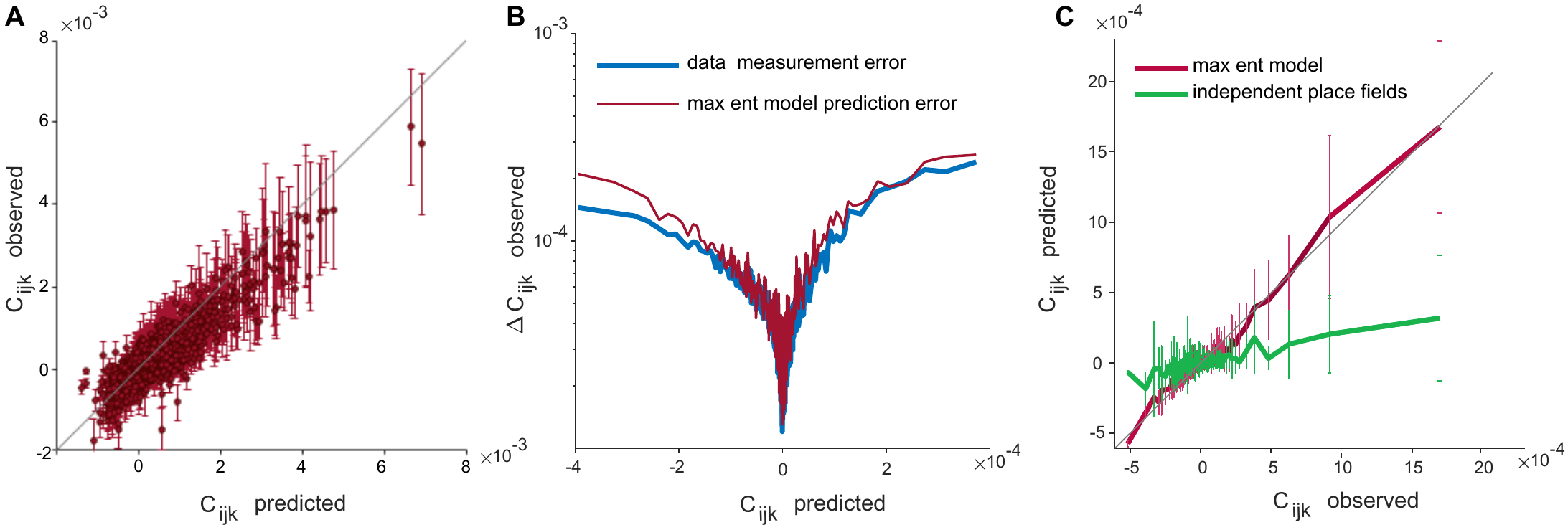}}
	\caption{\textbf{Triplet correlations.} \textbf{(A)} Predicted vs observed triplet correlations, $C_{\rm ijk}$ as defined in Eq (\ref{C3}). \textbf{(B)} Comparison of the maximum entropy model prediction errors for individual triplet correlations, and the measurements errors computed from the data itself. The two sets of errors are on the same scale, and are especially similar for the more common, smaller correlation values. \textbf{(C)} Predictions of triplet correlations from our maximum entropy model (red) and from an independent place field model (green). X axis is binned such that values represent a small range of 3-cell correlations rather than individual ones. The animal's location can only poorly account for the higher--order structure in the data. The maximum entropy model captures the higher order structure much better.\label{c3_all}}
\end{figure*}  

The model that we have defined predicts the probability of every possible state that the network could occupy. It is natural to think not about the probabilities themselves, but about their (negative) logarithm, which measures how surprised we should be by each particular pattern of activity \cite{Shannon1948}.  In the equivalent Boltzmann distribution, the negative logarithm of the probability is the energy.  The model makes a prediction for the distribution of energies, but we can also walk through the data, compute the energy of every state that we see, and thus estimate the distribution of energies in the data.  While states with more cells active are less likely (Fig \ref{test1}A), there are more ways of choosing these active cells out of the population, leading to a non--trivial distribution of energies, shown in Fig \ref{test1}B.  We see that theory and experiment agree very well out to energies $E \sim 25$.  This energy is equivalent to a probability of $e^{-E} \sim 10^{-11}$, corresponding to events that should occur once every few hundred years.  While of course we can't know if we have correctly predicted this tiny probability for any single state, what  we are seeing is that the model correctly predicts the number of these rare states, each of which is seen only once in the data.

Since we build our model out of the pairwise correlations among neurons, it is natural to test the predictions for higher--order correlations. The probability that $K$ out of $N$ neurons are active, as in Fig \ref{test1}A, is one combination of all the higher--order correlations, but we would like to look in more detail. To do this, we check the correlations among triplets of neurons,
\begin{equation}
C_{\rm ijk}\equiv \langle (\sigma_{\rm i} -\langle\sigma_{\rm i}\rangle) (\sigma_{\rm j}-\langle\sigma_{\rm j}\rangle)(\sigma_{\rm k}-\langle\sigma_{\rm k}\rangle)\rangle
\label{C3}
\end{equation}
These correlations are small, but significantly non--zero for most of the $7.6\times 10^4$ distinct triplets in the population. Figure \ref{c3_all}A shows that the model correctly predicts   these correlations, within error bars, across the full dynamic range of the data. 

To examine the quality of our predictions more closely, we zoom in on the small ($|C_{\rm ijk}|<4\times 10^{-4}$) correlations, and compare the error in our predictions with the experimental error in measuring these triplet correlations.  Concretely, we make small bins along the $C_{\rm ijk}$ axis, and within each bin we compute the mean square difference between predicted and measured correlations, as well as the mean square error in our measurements. We then plot the square root of these quantities, which we can think of as error bars on predictions and measurements, respectively,  in Fig \ref{c3_all}B. We see that the two measures of error are essentially equal across a wide range of correlation values:  our model  predicts the triplet correlations with a precision that essentially matches the experimental error; one cannot do better.

For neurons in the hippocampus, one obvious source of correlations is the existence of place fields:  cells with overlapping place fields should have positive correlations in their activity, and cells with distant place fields must be anticorrelated. We certainly see something of this pattern in the data of Fig \ref{basics}C, and it tempting to think that all correlations could be understood in this way, in which case our success in predicting triplet correlation may be less surprising.  To explore this possibility, we constructed a model in which cells respond independently to the position of the animal along the virtual track. Given  the probability of each cell being active as a function of position,  as in Fig \ref{data2}D, the model of ``independent place cells'' has no free parameters (see Appendix \ref{IPFmodel}), and in particular we can predict correlations among pairs or triplets of neurons. The prediction of individual pairwise correlations based on place fields alone is not bad (Fig \ref{C3place} in Appendix \ref{IPFmodel}), but the prediction of global properties of these correlations, such as their eigenvalues spectrum, is worse, and the prediction of triplet correlations is much worse. 

In Figure \ref{c3_all}C, we group triplets of neurons into bins based on their observed correlation, and plot the mean and standard deviation of predictions in each bin from both the maximum entropy model and the independent place field model.  For the maximum entropy model, this is just a different way of looking at the results in Fig \ref{c3_all}A, and correspondingly we see that predicted correlations form a tight band around the observed correlations.  But for the independent place cell model, predicted correlations span a much smaller dynamic range, and for the bulk of triplets with small $|C_{\rm ijk}|$ there is almost no systematic relationship between predictions and observations.  These results suggest, strongly, that correlations in the network are not simply a response to the environment, but are most accurately described as  a collective property of the network itself.

We further investigated if this collective behavior can be accounted for by a few dominant components that might exist in the network beyond place encoding. Among the natural candidates are a global network excitation signal and instability in place coding. As reported in Appendix \ref{collective} and shown in Fig \ref{low_rank}, these cannot account for the correlation structure the network exhibits beyond place encoding. It seems then that this population level property is not a result of any simple, low dimensional signal that we expect to see in the hippocampal network, and that it stems from both place and the non-place cells. 

If activity in a network is collective, then we should be able to predict the activity of each individual neuron  from the pattern of activity in the rest of the network. As noted above, the model we are exploring is mathematically equivalent to model for spins in a magnet, and thus the influence of the network on a single neuron can be summarized by an ``effective field''
\begin{eqnarray}
h^{\rm eff}_{\rm i} &=& E ( \sigma_1, \, \cdots ,\,\sigma_{\rm i}=0, \,\cdots , \,{\sigma_N} ) 
\nonumber \\
&& 
\,\,\,\,\,\,\,\,\,\, 
- E ( \sigma_1, \, \cdots ,\,\sigma_{\rm i}=1, \,\cdots , \,{\sigma_N} )\\
% \\
&=& h_{\rm i} + \sum_{{\rm j}\neq{\rm i}} J_{\rm ij}\sigma_{\rm j} .
\label{heff_def}
\end{eqnarray}
Positive effective fields favor neurons being active; more precisely  the probability of the single neuron $\rm i$ being active is given by the logistic function,
\begin{equation}
P(\sigma_{\rm i} =1| h^{\rm eff}_{\rm i} ) = \frac{1}{1+\exp(-h^{\rm eff}_{\rm i})} .
\label{logit}
\end{equation}
Thus, the model predicts the probability of any single neuron being active at any moment in time, given the state of all the other neurons in the network.

\begin{SCfigure*} 

\includegraphics{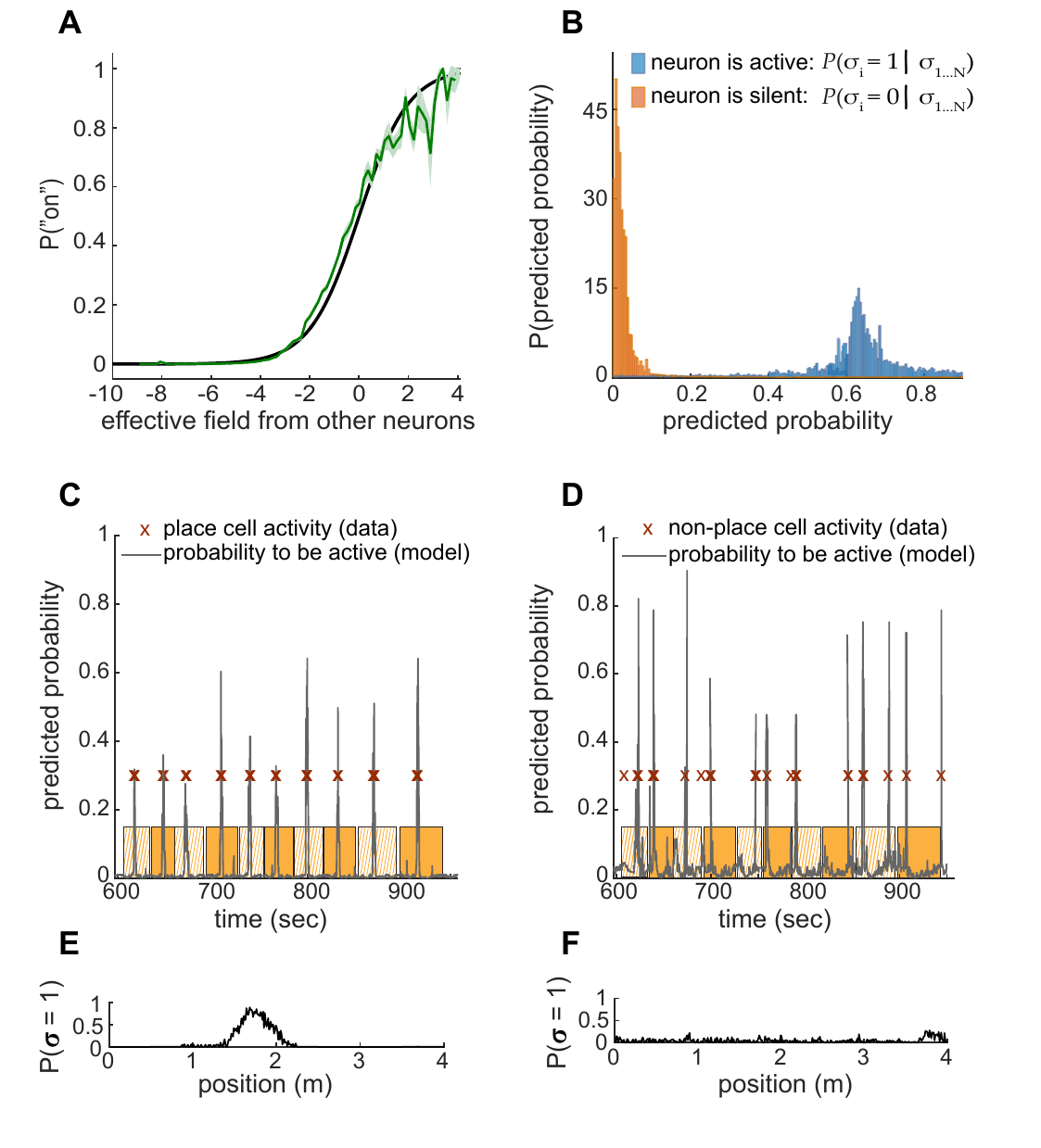}
\vspace{-20 pt}
\caption{\textbf{Population effective field predicts the activity of individual neurons.} \textbf{(A)} Probability of neuron to be active based on effective field. The relationship between the computed effective field and the probability of a neuron being active (green) compared with the parameter free prediction in Eq (\ref{heff_def}) (black). Shaded silhouette is standard deviation across. \textbf{(B)} Predicted conditional probability distributions for all neurons to be active/silent for all time points. $P(\sigma_{\rm{i}}=1|h_{\rm{eff}})$: time points where the neuron is active in the data colored in orange. $P(\sigma_{\rm{i}}=0|h_{\rm{eff}})$: time points were the neurons are inactive colored in blue. \textbf{(C)} Predicted probability vs time for a place cell. Red x marks show times where the neuron was active. The start and end of each run along the virtual track are drawn in yellow and striped yellow rectangles. \textbf{(D)} Predicted probabilities vs time as in panel C, but for a non--place cell. Example of all time points in a specific time window. Red x marks show time points where the neuron was active in the data. The start and end of each trial are drawn in yellow and striped yellow rectangles. \textbf{(E)} Probability of activity vs position for the place cell depicted above in panel C. \textbf{(F)} Probability of activity vs position for the non-–place cell depicted above in panel D. \label{heff_figs}}

\end{SCfigure*}

To test the predictions of the model, we walk through the data and at each moment, for each neuron, we compute the effective field given the state of all the other neurons [Eq (\ref{heff_def})]; we also mark whether the target neuron was active. Pooling all these data, we can plot the actual probability of a cell being active as a function of its effective field, and compare with the prediction of the model [Eq (\ref{logit})]. The results of this comparison, in Fig \ref{heff_figs}A, show that theory and experiment agree very well, across the full dynamic range. Additionally, if the effective field correctly predicts the probability of a neuron being active, then, conversely, a neuron being active should predict a high value of the effective field. Thus, if we compare moments when cell is active or inactive, we should see very different distributions of the predicted probability, and this is shown in Fig \ref{heff_figs}B. 

In Figures \ref{heff_figs}A and B, we have pooled probabilities for individual cells across time; in Figs \ref{heff_figs}C and D we focus on two particular cells (indexed 14 and 77 in Fig \ref{data2}).  We plot, as a function of time, the probability that a single neuron should be active, given the state of all the other neurons in the population; we emphasize again that the model makes an unambiguous prediction, and that there are no free parameters to adjust. The predicted probability consists of a series of relatively brief peaks, separated by periods of probability near zero, and this is reproduced in the data, where we see that these neurons are active only during the predicted peaks. The neuron in Fig \ref{heff_figs}C has a clear place field in the environment (Fig \ref{heff_figs}E), and the peaks of activity correspond to times when the mouse passes (virtually) through the place field. Importantly, the model we have built makes no reference to the position of the mouse.  Thus, to the extent that we are able to predict the activity of place cells, this is because information about place is represented collectively in the network. In contrast, the activity of the neuron shown in Fig \ref{heff_figs}D is hardly spatially modulated in this environment, yet the model still predicts brief periods of high probability for the cell being active, separated by longer periods of near zero probability, and again the actual moments of activity follow these predictions quite well. Thus, it appears that both place cells and the remaining neurons that do not code for a particular location, are part of the same patterns of collective activity. We emphasize that these results hold true independent of the specific definition of which neuron is a place cell and which is not. Our model predictions remain valid for any level of location encoding of the cell, whether significantly, partially, or hardly at all.

\begin{figure*} %{r}{1\textwidth}
	\begin{center}
		%  \vspace{-21pt}  
		\includegraphics[width=\linewidth]{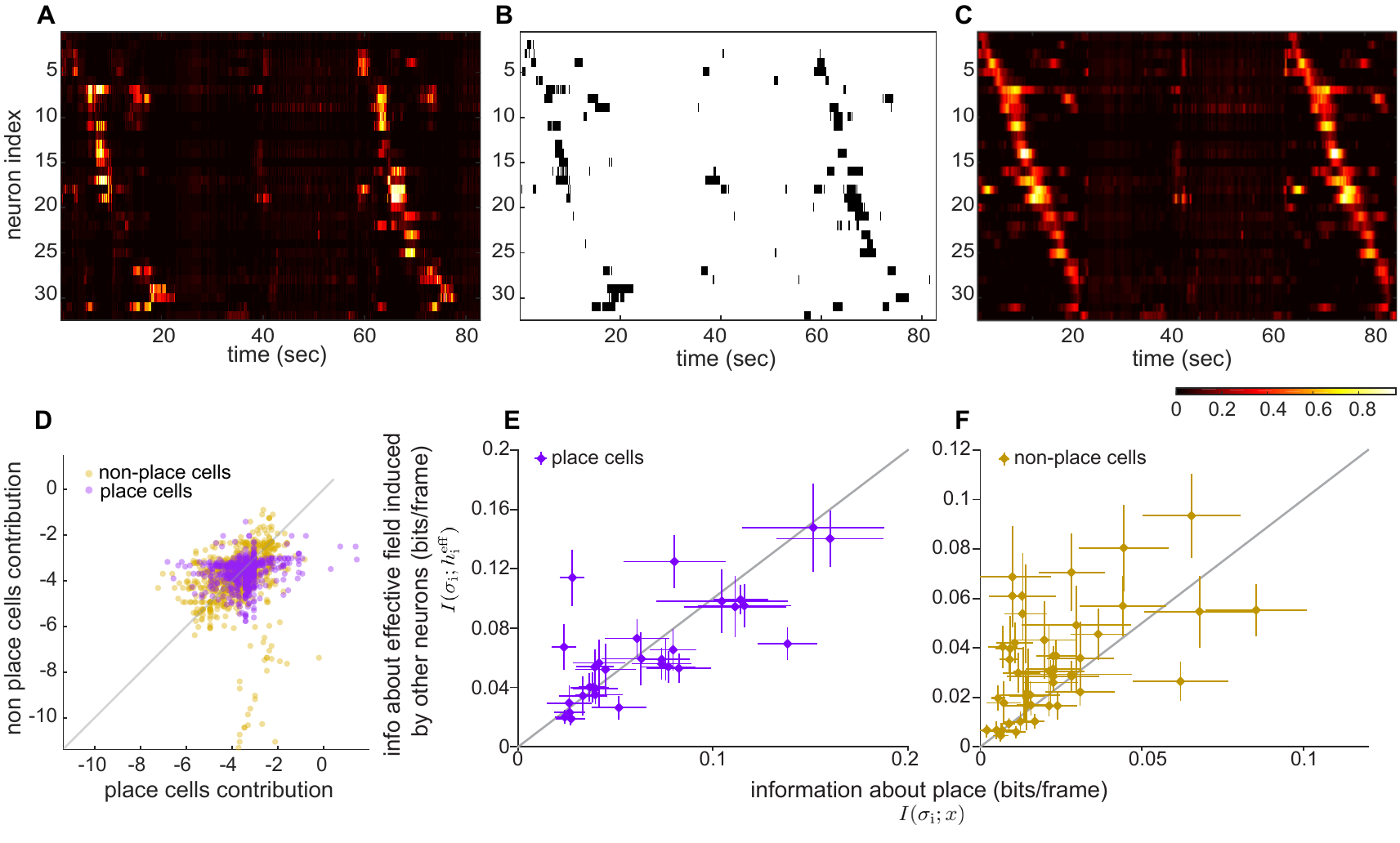}
	\end{center}
	% \vspace{-9pt}
	\caption{\textbf{Collective and place information} \textbf{(A)} Predicted probability computed from the effective field in our model for all place cells along two consecutive runs along the linear track. Time window shown is the same as in panels B and C. During the first run, cells 21-25 are predicted to ``miss'' their place field. Indeed, comparing to the real data shown in panel B, there is a missed field for these cells on that specific run, but all cells are predicted to be active in the second run. \textbf{(B)} Real data of place cells activity during two runs down the linear track, corresponding to panels A and C; note the ``missed'' events for cells 21-25 in the first run. \textbf{(C)} Predicted probability computed from the independent place cells model for all place cells along two consecutive runs along the linear track, as in A. Prediction for the two runs are almost identical, with no indication of when fields should be missed. \textbf{(D)} Contributions from place cells and non--place cells to the effective field. Each point shown is a time point from a specific neuron (sub--sampled 1:1000). The predicted probability includes a contribution from place cells (x axis) and from non-place cells (y axis). A point was colored purple if the time point belonged to a place cell, and yellow if the time point belonged to a non-place cell. Network contributions to both place and non--place cells seem to originate equally from the two sub--populations.  (E) Each point (with error bars for both axis) corresponds to the information that the activity of a particular place cell carries about the state of the network vs. the information carried about the position of the animal. Place cells carry similar amount of information about the state of the network and about the animal's position. \textbf{(F)} As in panel E, but for non--place cells. Non--place cells carry more information about the state of the network than about the position of the animal. Note that some of them carry non--negligible position information, even though they do not have a sufficiently reliable localized place field to be classified as place cells. \label{heff_2} } 
\end{figure*}

In addition to treating all neurons in the population on an equal footing, we also looked into moments where place cells fail to exhibit their typical place-modulated activity. In Figure \ref{heff_2}, panel A shows the probability predicted from our model for each of the the place cells in the population during two successive runs along the virtual track. Panel B shows for comparison the neural activity for the same two runs. The first sequence is an example for a run with “missed” fields (neurons 21 to 25) while in the second sequence all place cells are active at the expected times. We see that our model of collective activity in the network manages to capture accurately even these moments where a place cell misses its place field. For the second run where all neurons were active, the model predicts no missed fields. In contrast to the success of our model for collective activity, treating the cells as having independent place fields necessarily predicts the same pattern of activity during every run along the track (Fig \ref{heff_2}C). Any variations must be ascribed to “noise” in the system. In contrast, we find that a cell “missing” its place field is not an individual deviation disconnected from the rest of the population, but rather that the network state is predictive of the place field being dropped.  

The effective field $h_{\rm i}^{\rm eff}$ [Eq (\ref{heff_def})], which determines the probability of a single neuron being active, can be thought of as having three components.  First, each neuron has an intrinsic bias $h_{\rm i}$, independent of all the other cells in the network.  Activity in other neurons adds to this bias, through the terms $\propto J_{\rm ij}$ in Eq (\ref{heff_def}), and we can further separate this sum into contributions from place cells and from non-place cells, which are the second and third components of the effective field. In Fig \ref{heff_2}D we show a scatter plot of these last two contributions, separating the case where the target cell i is itself a place cell or a non-place cells. Perhaps the most important conclusion is that contributions from the two classes of cells are comparable: our ability to predict the activity of one cell from the patterns of activity in the rest of the network depends both on neurons with a clear place field in the environment, as well as on those that are not strongly place modulated, more or less equally. This reinforces the conclusion that regardless of the level of place modulation in their activity, all cells are part of the same patterns of collective activity.

\section{Discussion}

We have explored a description of activity in the hippocampus that focuses on the collective behavior of the network, rather than on the sensory inputs or motor outputs.  Extending a strategy that was originally applied to populations of neurons recorded in the retina – a “lower” sensory area \cite{Schneidman2006,Tkacik2010,Ganmor2011,Tkacik2014}, we have built this model by taking from experiment a few basic quantities, in particular the mean activity of each neuron and the correlations between activity in pairs of neurons.  Beyond these constraints from experiment, we ask for a model that generates states of activity in the network that are as random as possible, so that the only structure is that required to match the constraints. This maximum entropy model is learnable from the data; given the size of the data sets to which we have access, there are no signs of over-fitting (Figure \ref{all} in Appendix \ref{more_expts}). The model is simple, yet it passes a number of quite detailed tests.

Although the model is based on measurements of correlations between pairs of neurons, it makes predictions for the probability of any possible combination of activity and silence in the network.  While we cannot estimate the probabilities for each of these $2^{78} \sim 3\times 10^{23}$ states, we can measure the probability that $K$ out of the $N$ neurons in the network are active simultaneously (Fig \ref{test1}A), and we can measure the correlations among all  $7.6 \times 10^4$  distinct triplets of neurons (Fig \ref{c3_all}A), and in both cases, we see detailed, quantitative agreement between theory and experiment. The central quantity in our model is the “energy” or (negative) log probability, and we can predict the distribution of this quantity across the network states that actually occur in the experiment, deep into the tail of rare states (Fig \ref{test1}B). Finally, and perhaps most importantly, we can test the idea that activity in the network is collective by predicting the activity of one neuron based on the state of all the others (Fig \ref{heff_figs}).  We have focused on the analysis of a single data set, from 78 neurons in a half an hour recording session, but essentially the same results are found in 5 other data sets, as described in Appendix (Fig \ref{app_learning}).

Our first conclusion is that the general strategy we have adopted succeeds in describing the distribution of activity across networks of $\sim80$ pyramidal neurons. These results indicate that the maximum entropy method is a viable path for constructing global models of joint activity in central brain regions. We may yet find that the small quantitative discrepancies between theory and experiment are hints of things we are missing that become clearer in larger networks, as in the retina \cite{Ganmor2011a,Tkacik2014}, but the quantitative agreement that we do find here sets a standard for what we should expect from such models. We emphasize that the correlation structures we see in this network are not simple: covariation among neurons cannot be captured by linear projection into a low dimensional space (Fig \ref{heff_figs}), nor can the higher--order correlations be captured by a model of independent place cells, even allowing for arbitrarily complex and disconnected place fields (Fig \ref{c3_all}C and Appendix \ref{IPFmodel}). 

In the view of network activity as a collective whole, we think of individual cells as following the state of the network, as summarized by the ``effective field'' [Eq (\ref{heff_def})].  An alternative view is that each neuron is encoding information about the organism's behavior and its sensory inputs. In the hippocampus, the natural variable is the position of the mouse along the (virtual) track, and indeed 32/78 of the cells in the population that we study here have clear place fields (Fig \ref{data2}) but the majority of cells (46/78) do not.  These  ``non--place'' cells, whose activity is not significantly spatially modulated, could function as place cells in another environment, but they are far from being ``silent'' \cite{Thompson1989,FergusonJacksonRedish2011}.  Because our model does not make reference to place, both classes of neurons appear on the same footing, but this leaves open the question of what really is being represented.

The activity of individual place cells, by definition, carries information about position, and we can quantify this in bits: the information $I(\sigma_{\rm i};x)$ that the state of neuron $\rm i$ at one moment in time provides about the position of the mouse along the virtual track.  Since the cells are active only in a few percent of the time frames, we expect this average, momentary information to be small in absolute terms, roughly $\sim 0.1$ bits, as seen in Fig \ref{heff_2}E.  But we can also ask how much information the activity of these neurons provides about the effective field, that is a summary of the state of the rest of the network, $I(\sigma_{\rm i};h_{\rm i}^{\rm eff})$. Fig \ref{heff_2}E shows that, except for a few outliers, these two information measures are equal within error bars, for each place cell.

Non--place cells are defined by the fact that there is no single place near which the probability of being active is high. Nonetheless, this activity in these cells can convey information about place, although typically less than the place cells, as we expect.  For almost all of the non--place cells, however, we find that the information that activity provides about the effective field is much larger than the information about position, $I(\sigma_{\rm i} ; h_{\rm i}^{\rm eff} ) \gg I(\sigma_{\rm i} ; x)$, as shown in Fig \ref{heff_2}F. This means that the information shared among neurons in the network is not entirely about place. We suspect that it will be easier to discern the biological significance of this collectively encoded information when we can monitor larger populations of cells in a broader range of environments.

Our results show that the activity of both place and non--place cells can be given a compact, unified description as part of a single network exhibiting collective behavior.  All the quantitative features of this collective behavior are predicted from the observed correlations between pairs of cells.  We have seen that higher order correlations are not predicted at all in a model where cells respond independently to the animal's location (Fig \ref{c3_all}C and Fig \ref{C3place}C), and that there are even quantitative deviations of the pairwise correlations from this place field model (Fig \ref{low_rank}C in Appendix \ref{collective}). Neither the pairwise correlations $C_{\rm ij}$ in the data nor the inferred interactions $J_{\rm ij}$  in our model are of low rank, and the deviations of the correlations from the place field model also are not of low rank, suggesting that there is no global additive modulation of activity (Fig \ref{low_rank}D in Appendix \ref{collective}).  Place fields can shift from trial to trial \cite{Fenton1998}, but if we estimate these shifts from our data there is no obvious global pattern that could generate the appearance of collective behavior (Fig \ref{low_rank}D in Appendix \ref{collective}). Independent of its underlying mechanism, the fact that the network exhibits a nontrivial pattern of collective behavior that is captured by our relatively simple model does not seem to be an obvious consequence of known deviations from a simple place field model, and certainly such behaviors have not been predicted quantitatively by previous models.

In our analysis, we investigated equal--time correlations; a reasonable next step is to incorporate temporal correlations into the model. While recent studies have offered interesting adjustments to our framework to account for time dependencies to data \cite{mora2015dynamical}, the dynamical aspect of the collective behavior of place and non--place cells remains a challenge.  

An interesting phenomenon observed experimentally in dorsal hippocampus is ``remapping'', which occurs when an animal is moved from one familiar environment to another, and the place cells in CA1 code for a different cognitive map in each environment. When discussing possible models for the hippocampal network, it is important to remember that these different maps partially overlap --- a fact which has been difficult to resolve with the initial view of different spatial maps needing to be orthogonal to one another. This conflict has led to two possible mechanistic hypotheses: either we need a more sophisticated attractor map architecture than has been proposed previously, or the hippocampus conjunctively encodes both map information and some other type of information \cite{SkaggsMcNaughton1998}. Our finding that each neuron has access to the state of the rest of the network, and that individual neurons' activity, regardless of how well they code position, are very well predicted by the network, points toward the latter hypothesis. Taken together, the evidence that the correlation patterns in the data only partially arise from place encoding and that additional information is encoded in the population--level, lead us to believe that interactions among neurons in CA1 may have equal or even greater influence than the animal’s position on the observed neural activity.

\section{Experimental methods}

\textbf{Surgery.} Optical access to the hippocampus was obtained as described in \cite{Dombeck2010}. Briefly, a $\sim$3mm circular hole was cut into the skull centered at 1.8 mm lateral and 2.0 mm posterior to bregma on the left side of the skull. The cortical tissue overlying the hippocampus was aspirated, and a circular metal cannula with a \#1 coverslip affixed to the bottom was implanted, with a thin layer of Kwik--sil (WPI) between the hippocampus and coverslip.\\

\textbf{Virtual reality setup.} Water--restricted mice were trained to receive water rewards by running on a 4m virtual linear track using a virtual reality setup similar to that described in \cite{RickgauerDeisserothTank2014}.  Mice ran on a Styrofoam treadmill constrained to rotate only around the horizontal axis (pitch) to prevent turning.  Treadmill movement was read out using an optical mouse, and a visual display of the virtual environment was projected onto a 270$^{\circ}$ toroidal screen. Images were acquired using two photon laser scanning microscopy as previously described \cite{Dombeck2010}.\\

\textbf{Image processing.} Images were analyzed using toolboxes and custom scripts written in Matlab.  Acquired images were motion--corrected by aligning to a template image, and cell finding was performed on the corrected movie using a modified version of the PCA--ICA algorithm \cite{MukamelNimmerjahnSchnitzer2009}.

%\subsection*{Author contributions}
%	Modelling and analysis, L.M.; Data collection, J.L.G, Writing -- Original Draft, L.M. and W.B.; Writing -- Review and Editing, L.M., J.L.G, C.D.B, D.W.T. and W.B; Supervision, C.D.B, D.W.T. and W.B.; 

\begin{acknowledgements}
	We thank G Berman, T Mora, B Scott, G Tka\v{c}ik, and J Zhou for many helpful discussions. Work supported in part by the National Science Foundation (Grants PHY-1305525, PHY-1451171, CCF-0939370), the Simons Foundation and the Howard Hughes Medical Institute. L Meshulam is a Howard Hughes Medical Institute International Student Research fellow.
\end{acknowledgements}

\appendix

\section{From continuous fluorescence signal to a binary time series}
\label{cont_bin}
To convert the neural activity reported by the kinetics of the GCaMP3 indicator, we used independent exponential fits for the rise and decay time of the fluorescence signal for each activity event. After identifying all maxima that can potentially be activity events, as described in the main text, the rising phase of each response was fit with a single exponential function. Traces whose time course deviated more than 3σ from average were discarded, since they usually reflected residual motion artifacts or external noise sources. Next, the decay trace was fit to a double exponential $f(t)=A_0+A_1 e^{-k_1 t}+A_2 e^{-k_2 t}$. In the case that $k_2$ was slower than 1$\sigma$ from the average trace for that cell, the time point was assigned a value of $1$, as it was assumed the cell kept on firing and the decay slope was fit again from the next time point and on, repeating until the slope was within 1$\sigma$ from the average. In the case the trace decayed faster than 1σ from the average for that neuron, the time point was assigned a $0$ value.

\section{Learning the model}
\label{app_learning}

\begin{figure} %{r}{1\textwidth}
	\begin{center}
		%  \vspace{-21pt}  
		\includegraphics{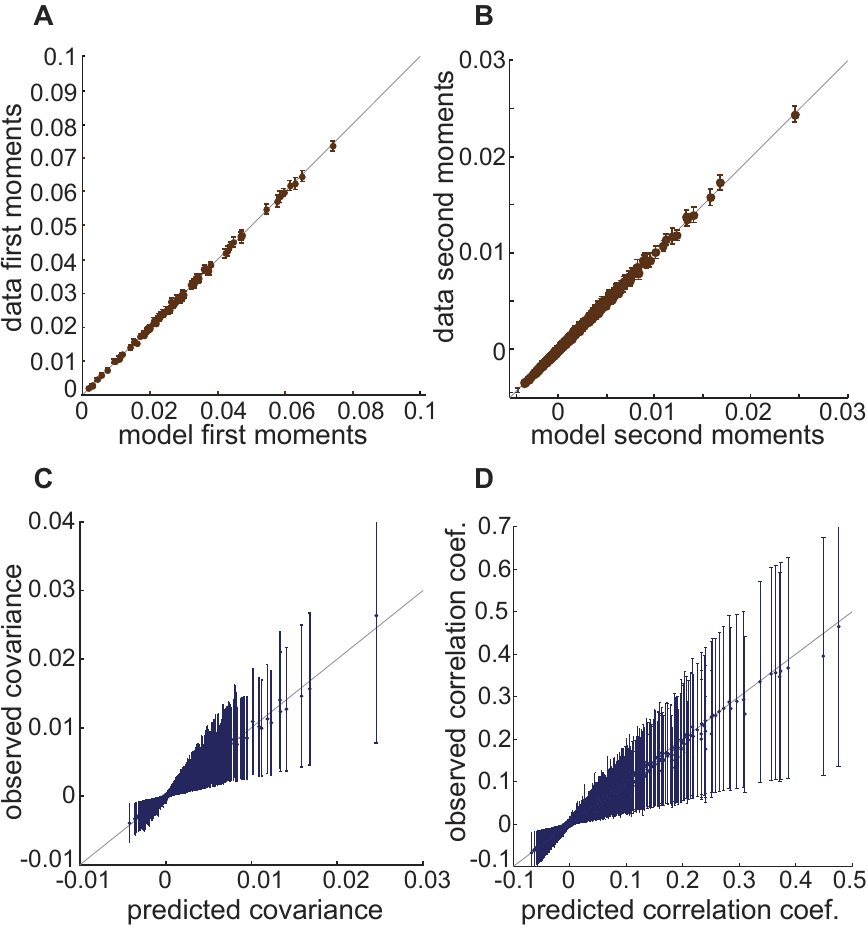}
	\end{center}
	% \vspace{-9pt}
	\caption{\textbf{Fitting the model –- comparing model vs. data}.  \textbf{(A)} First moments after the MCMC has converged. \textbf{(B)} Second moments, after the MCMC has converged. \textbf{(C)} $C_{\rm ij}$ pairwise covariances values as given by the model that was learned from ¾ of the data, compared to their values in the ¼ of the data that was not used to fit the model. Values are very well fit. \textbf{(D)} $c_{\rm ij}$, pairwise correlation coefficients as given by the model that was learned from ¾ of the data, compared to their values in the ¼ of the data that was not used to fit the model. Values are very well matched. \label{fit}}
\end{figure}

To build the maximum entropy models constrained by the firing rates and covariances computed from the data, we performed Markov Chain Monte Carlo (MCMC). In every cycle, we shift the values of our model parameters $\{h_{\rm {i}}, J_{\rm {ij}}\}$ in the direction of the gradient towards the solution. The log--linear form of the Boltzmann distribution guarantees us convergence for any set of observed parameters $\{h_{\rm {i}}, J_{\rm {ij}}\}$. The algorithm we use leverages quasi--Newton approximations, which estimate the Hessian of the target distribution from previous samples and gradients generated by the sampler \cite{schmidt2012ugm}. To ensure our final solution is not sensitive to initial conditions, we incorporated a long enough burn--in period (usually 100,000 cycles) where we threw away the samples before starting to record them. The estimate for our model parameters $\{h_{\rm {i}}, J_{\rm {ij}}\}$, which serve as the coefficients of the observables in our cost function, lets us obtain the full probability distribution for the joint activity of the neurons. Once we have the distribution, we can sample synthetic data from it, i.e. simulate a set of configurations (population states), according to their probability to occur in our model. Our probability distribution assigns a probability value for any possible configuration. To text for overfitting, we used $3/4$ of trials (randomly selected) to train the model, and left out a $1/4$ as a test set. Figure \ref{fit} shows the correlation values predicted by the model as computed from the sampled population states, against the ones computed from the test set. We can see they are in agreement within error bars.

\section{Survey of additional experiments}
\label{more_expts}

\begin{figure} [b]%{r}{1\textwidth}
	\begin{center}
		%  \vspace{-21pt}  
		\includegraphics{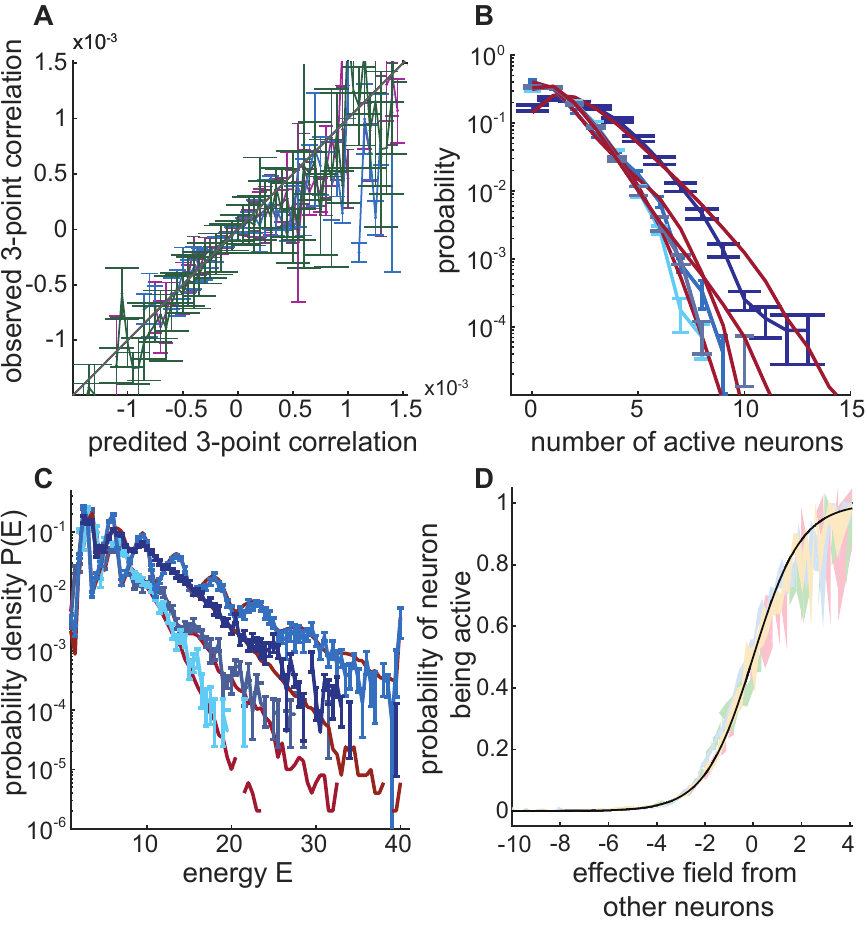}
	\end{center}
	% \vspace{-9pt}
	\caption{\textbf{Summary of 4 additional datasets} \textbf{(A)} Triplet correlations for 4 datasets. Corresponding to the red line in Fig 6C. \textbf{(B)} Probability vs number of active neurons. Corresponding to Fig \ref{steps}A. \textbf{(C)} Energy vs probability structure. Corresponding to Fig \ref{test1}B. \textbf{(D)} Probability of neuron to be active, overlaid on the logit function. Corresponding to Fig \ref{heff_figs}A. \label{all}}
\end{figure}

We analyzed 6 data sets recorded in CA1 of 4 GCaMP3 transgenic mice. The number of neurons in the data sets ranged 21-78: 1. $N=78$ (as described in the main text), 2. $N=69$, 3. $N=21$, 4. $N=68$, 5. $N=46$, 6. $N=40$. Each session lasted $T\sim 25$ minutes, where each run along the track lasted $K\sim 15$ sec. A separate model was constructed for each session.  Performance of the model for these datasets is summarized in Fig \ref{all}. Four of the datasets are shown for ease of visualization.

\section{Independent place cells}
\label{IPFmodel}

\begin{figure*}
	\includegraphics[width=\linewidth]{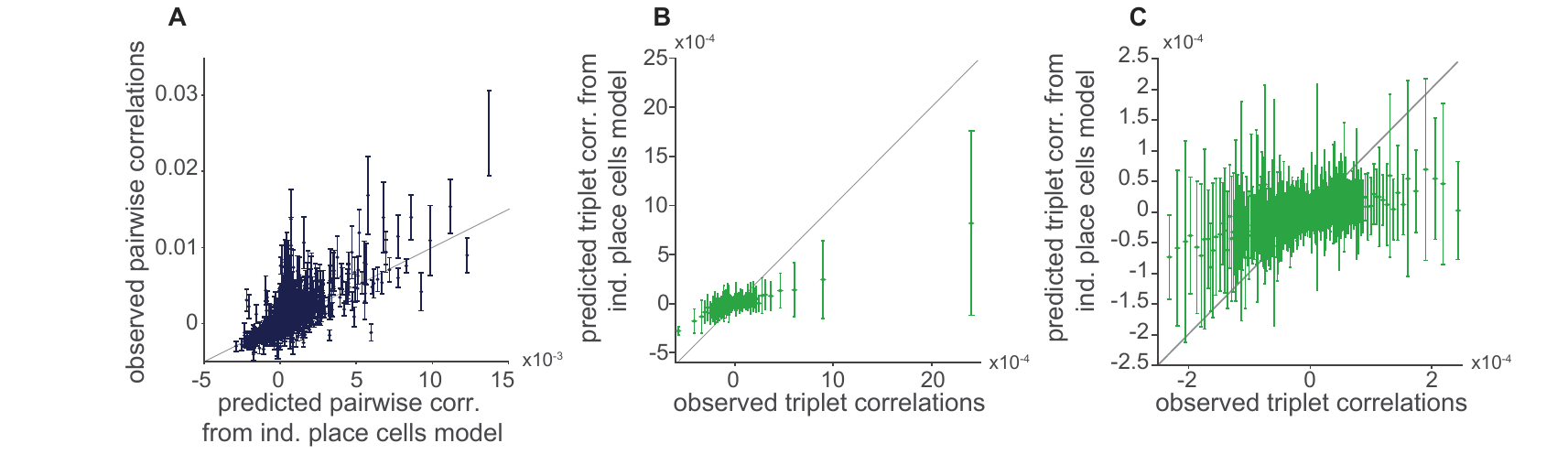}
	\caption{{\bf Pairwise and triplet correlations from the independent place cell model.} \textbf{(A)} Observed pairwise correlation vs the predicted pairwise correlations from the independent place cells model. Many of the values are well predicted, though some are in significant disagreement with the observed values. \textbf{(B)} Predicted triplet correlations from the independent place cells model vs the observed pairwise correlation. (This is a detailed version of the green line in Fig \ref{c3_all}C in the main text). The independent place cells model fails to predict the third order correlations in the data. \textbf{(C)} Zoom in on the most common, small correlations depicted in panel B. \label{C3place}}
\end{figure*}

We have explored a model in which the states of neural activity are a coherent, collective whole, with each cell's probability of being active determined by the activity of the other cells in the network.  This model makes no explicit references to external inputs, although these may be encoded implicitly through the pattern of correlations.  A very different approach is to imagine that each neuron is ``responding'' independently to the relevant sensory stimuli, and since we are in the hippocampus the natural stimulus variable is the (virtual) location of the mouse.  Indeed we made use of intuition from this model when we rationalized the pattern of pairwise correlations, arguing that the negative correlations arise because cells representing different places in the environment cannot be active simultaneously.

The idea of tracing correlations in neural activity back to correlations in their inputs is very old, dating back to a time when it was hoped that analysis of correlations could provide an unambiguous measure of anatomical connectivity \cite{MoorePerkelSegundo1966, Bullock1970}.  The decomposition of correlations into stimulus--driven (``signal'')  and intrinsic (``noise'') components remains popular in the analysis of more peripheral sensory systems. The standard approach in those cases is to present precise repetitions of the sensory input, and compare the correlations among neurons computed from properly aligned  with those computed after the trials have been shuffled, leaving only the stimulus--driven correlations.  Importantly, this is an experiment we can perform as observers of the brain, but not a distinction that is available to the brain itself.  Nonetheless it is interesting to ask how model of neurons independently encoding place would compare to the maximum entropy model of collective activity that we have considered here.

Following conventional methods, we can estimate the probability that each neuron will be in the ``on'' state during the time bin when the mouse visits a particular place along the virtual track as in Fig \ref{data2}C We see that, for 32/78 neurons, there is a clear position near which the probability of being active is very high, even close to one, with activity having near zero probability away from this point; these are classical place fields.  For the there 46/78 cells, however, the peak probability of being active is not very high, the activity is much more diffuse, and in several cases there are multiple low peaks.  We have referred to these as ``non--place cells'' in the main text, although it is clear that they carry some spatial information, even if not conforming to the classical picture of place cells.

Let us refer to the place field of cell $\rm i$ as $F_{\rm i}(x)$; this is the probability that cell $\rm i$ will be active in a time bin when the mouse is at virtual position $x$. If all cells are responding to the position variable independently, then we can generate synthetic sequences of activity just by reference to these probabilities as they vary along the trajectory $x(t)$.  Further, we can predict the expectation values of the binary variables $\{\sigma_{\rm i}\}$ and their correlations:

\begin{eqnarray}
\langle \sigma_{\rm i}\rangle_{\rm pl} &=& \int dx\,P(x) F_{\rm i}(x),\\
\langle \sigma_{\rm i}\sigma_{\rm j}\rangle_{\rm pl}  &=& \int dx\,P(x) F_{\rm i}(x)F_{\rm j}(x),\\
\langle \sigma_{\rm i}\sigma_{\rm j}\sigma_{\rm k}\rangle_{\rm pl}
&=& \int dx\,P(x) F_{\rm i}(x)F_{\rm j}(x)F_{\rm k}(x) ,
\end{eqnarray}
where the subscript $\rm pl$ reminds us that these predictions arise from a place field model.
In Fig \ref{C3place}A we show the predictions for the pairwise correlations, 
\begin{equation}
C_{\rm ij}^{\rm pl} = \langle \sigma_{\rm i}\sigma_{\rm j}\rangle_{\rm pl} -
\langle \sigma_{\rm i}\rangle_{\rm pl}\langle \sigma_{\rm j}\rangle_{\rm pl} ,
\end{equation}
compared with what we see experimentally.

The predictions from an independent place field model do a reasonable job of reproducing the observed pairwise correlations.  Although one can find significant disagreements, the overall trends are captured, and there is even detailed quantitative agreement between model and data for many of the pairs.  The picture changes substantially when we try to predict the correlations among triplets of neurons, as we did with the maximum entropy model in Figs \ref{test1}A and B. In Fig \ref{C3place}B we show the triplet correlations predicted by the model of independent place fields, compared with the experimental data. 

The model of independent place fields significantly underestimates the correlations among triplets of neurons, so that a plot of predicted vs observed $C_{\rm ijk}$ has a slap of $\sim 1/2$ across the full dynamic range (Fig \ref{C3place}B).  But if we zoom in on the most common, small correlations, we see that the situation is much worse. 
Throughout the window shown in Fig \ref{C3place}C, the model of independent place field predicts essentially zero triplet correlations,   

\section{Defining place cells}

\label{def_place_cells}

In order to identify which neurons qualify as place cells, we first divided the virtual 4-meter-long track into 20 spatial bins and summed the number of “on” moments every neuron had in each bin. Following the strategy described in \cite{mcnaughtoninformation}, to categorize how good a description of a location is given by an individual neuron, we measured the extent to which the information is concentrated in a specific place. We assigned 5 spatial bins as the range for a potential place field. The neuron’s mean activity (over all trials) center of mass determined the location field’s center bin. Spatial bins of the place field were defined as the center bin and 2 more bins from each side. Next, we implemented a two--step approach as described in previous studies. First, we computed the spatial information for each neuron per place field \cite{mcnaughtoninformation,yartsev2011grid,domnisoru2013membrane}.
\begin{equation}
I = \sum_{k=1}^{5} p_k \frac{\langle\sigma_{\rm i}\rangle_k }{\langle\sigma_{\rm i}\rangle}\log_2\frac{\langle\sigma_{\rm i}\rangle_k }{\langle\sigma_{\rm i}\rangle} .
\label{def_place}
\end{equation}

where $p_k$  is the probability of the mouse to be in the $k$-th bin of the place field, $\langle\sigma_{\rm i}\rangle_k$ is the mean activity of neuron $\rm i$  in the $k$-th bin of the place field, $\langle\sigma_{\rm i}\rangle$ is the overall mean activity and $k$ runs over the spatial bins of the place field. Note that if we were to extend this calculation to all bins, we would get the measure for any place information in the neural signal (not necessarily localized), $I(\sigma_{\rm i};x)$, described in the main text. The second step in our two--step approach is to perform a shuffling test \cite{langston2010development,wills2010development}, shifting the entire sequence of activity for every neuron in each trial by a random interval. The end of the run was wrapped to the start to create a circular permutation, and the interval was picked at random from a uniform distribution of all durations between 3 secs to the trial’s duration. We repeated the process 500 times per neuron and computed the spatial information each time. This procedure has the advantage of effectively decoupling the neural activity from the mouse’s location, while not breaking its temporal structure. A neuron qualified as a place cell if its information per active time bin (``on'' moment) was at least in the 80th percentile of the shuffled distribution for that neuron. In the case of a neuron with more than one field, the procedure was performed for each field separately and success in either of them would have qualified the neuron to be a place cells (rare).  Out of the 78 neurons in the main data set we reported here, 32 qualified as place cells according to the shuffling criterion. This fraction was similar in the other 5 datasets we analyzed, ranging from 28\% to 41\%.

\begin{figure} %{r}{1\textwidth}
	
	\includegraphics{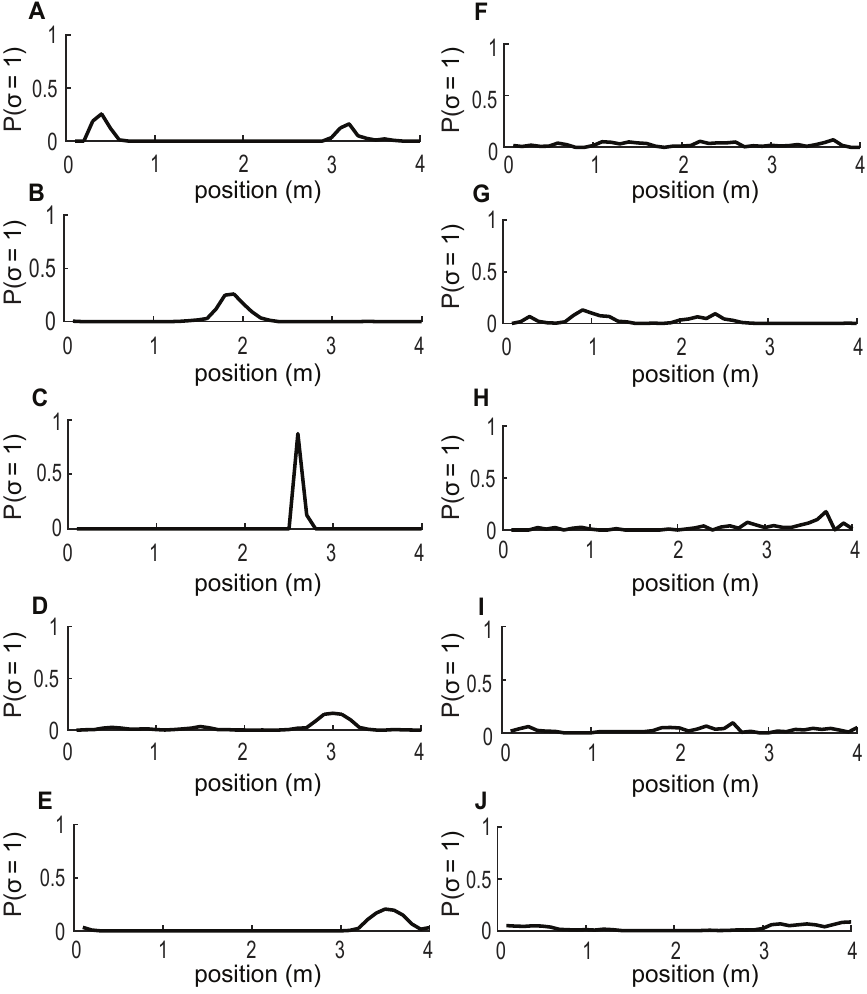}

	\caption {{\bf Place fields of place and non-–place cells.} \textbf{(Left column)} place fields of randomly chosen place cells out of the top 32 cells reported in Fig \ref{data2}D in the main text. Panels A-E correspond to cells \# 2, 8, 15, 26, 30. Cell \#2, shown in panel A, is an example for a rare case of a place cell with two place fields. \textbf{(Right column)} place fields of randomly chosen cells that did not pass the criterion to qualify as place cells out of the bottom 46 cells reported in \ref{data2}D in the main text. Panels F-J correspond to cells \# 38, 42, 51, 61, 75. Note that some of them do have a certain level of spatial tuning. \label{examps}}
	
\end{figure}

\section{Investigating the collective behavior of the network }
\label{collective}

\begin{figure*} %{r}{1\textwidth}
	\begin{center}
		\includegraphics{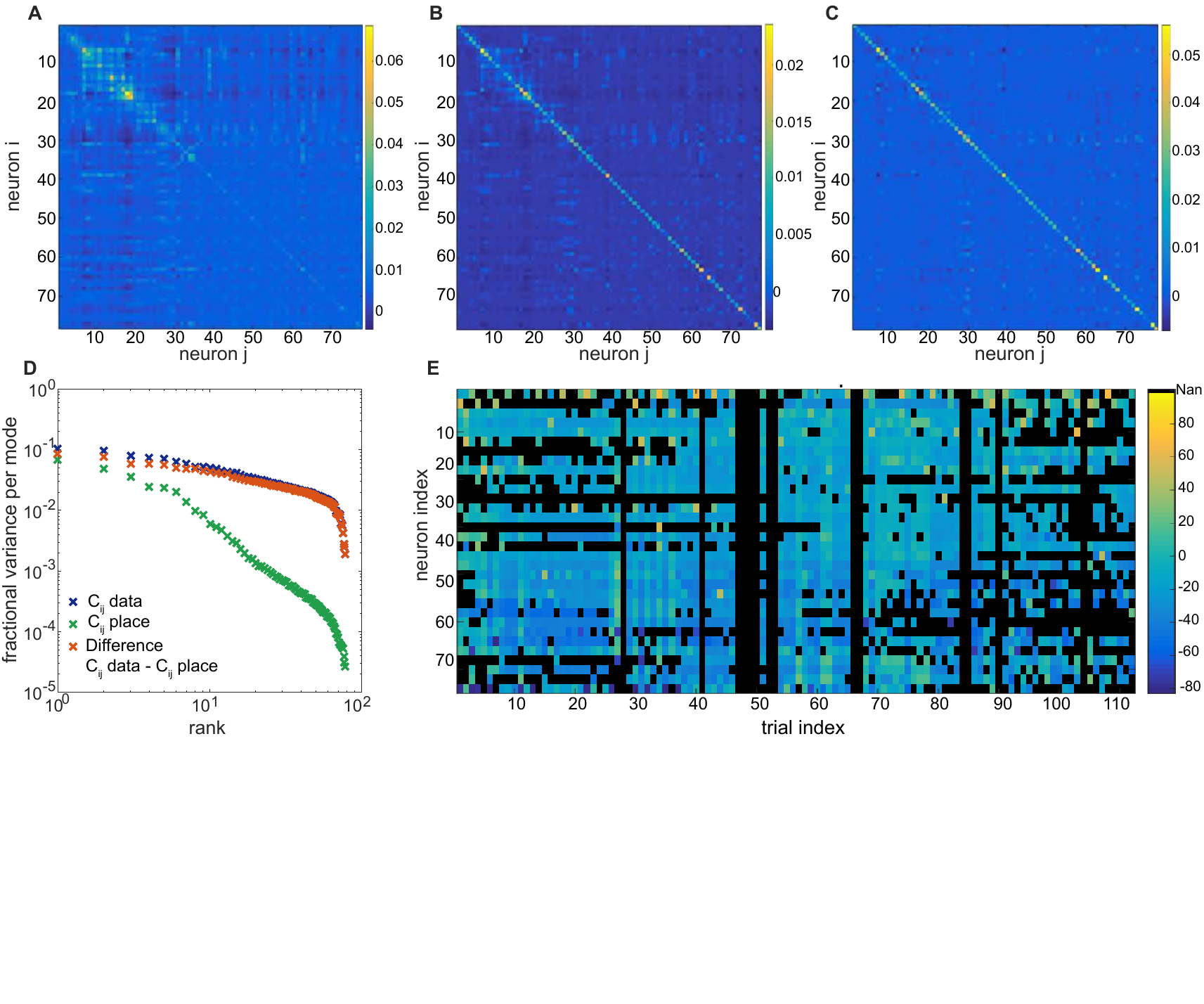}
	\end{center}
	\vspace{-110pt}

	\caption{\textbf{Alternative explanations for the collective behavior of the network.} \textbf{(A)} Pairwise covariance matrix $C_{\rm ij}$  of the data. \textbf{(B)} Pairwise covariance matrix $C_{\rm ij}$  of the activity that is accounted for by place--related information (i.e. independent place fields model). \textbf{(C)} Pairwise covariance matrix $C_{\rm ij}$ of the difference between the covariance matrix in panel A and the covariance matrix in panel B. These are the correlations of any part of the activity that cannot be explained by information about locations. \textbf{(D)} Probability of neuron to be active, overlayed on the logit function. \textbf{(E)} Trial--by--trial variability in place fields locations. Colors represent $\delta L_{\rm il}$ as defined in Eq (\ref{shift_def}), the change in the center of mass of the activity per neuron per trial, relative to its mean location over all trials (``real'' center). NaN values indicate trials where the field was missed or mean activity was too low to determine a meaningful center of mass. No single global mode captures all the variance. \label{low_rank}}
\end{figure*}

To further explore potential explanations that could account for the covariation of the neurons that we observe, since place--coding alone does not, we looked into two known sources of variability in CA1 neural activity: global changes in the network excitation and trial-by-trial changes in place field locations. 
First, we ask whether the global pattern of correlations among pairs of neuron (Fig \ref{low_rank}A) might be the result of activity in the network being confined to a space with dimensionality much lower than the number of neurons, as has been suggested in other systems \cite{mazor2005transient,machens2010functional,mante2013context}. To test this possibility, we compute the eigenvalues of the covariance matrix $C_{\rm ij}$ as defined in Eq (\ref{Cij_def}) in the main text, with the results shown in blue in Fig \ref{low_rank}D. The covariation among neurons cannot be captured by linear projection into a low dimensional space, nor can the higher--order correlations be captured by a model of independent place cells, even allowing for arbitrarily complex and disconnected place fields (Fig \ref{C3place}C and Fig \ref{c3_all}C in the main text). Moreover, the eigenvalue spectrum of the difference in correlations between the data and the place--related activity shows no sign of being low rank either (Fig \ref{low_rank}D (in red)), indicating that no simple model for global modulation of the network that we might attempt to write down could successfully capture it.

Next, we turn to the trial-by-trial changes in place field locations. This instability of the place cells’ code, originally referred to as ``excess firing variance'' \cite{Fenton1998},  is characterized by changes in discharge during different passes of the animal through the same location. In the positional domain, place fields occasionally get dropped or even missed completely (e.g. Fig \ref{heff_2}B in the main text). We explored the changes in their locations by comparing trials. In Fig \ref{low_rank}E we show all place fields location changes, $\delta L_{\rm il}$: 
\begin{eqnarray}
x_{\rm i}^{CM} &=&  \frac{\sum_t x(t)\sigma_{\rm i}(t)}{\sum_t\sigma_{\rm i}(t)} \\
\delta L_{\rm il} &=& x_{\rm il}^{CM}\langle x_i^{CM} \rangle 
\label{shift_def}
\end{eqnarray}

For neuron $\rm i$  in any trial $\rm l$, $x_i^{CM}$ is the place cell’s activity’s center of mass in that trial, and $\langle x_i^{CM}\rangle$  is the center of mass averaged across all trials (``true location''). Decomposition of the location shifts’ matrix (Fig \ref{low_rank}E) reveals no single pattern or global mode that could account for the collective behavior of the system. 
We conclude that neither of the global phenomena we tested accounts for the collective nature of the system. Indeed, we have not ruled out the possibility that there is a different variable of a more complex nature that is nonlinearly embedded in the data and could capture the observed covariation beyond position coding. However, since neither the covariance matrix of the data itself, nor the difference between this matrix to the covariance matrix of place--related activity only, is low rank, no single linear variable could help explain the data simply by addition to the low dimensional feature of place coding.   

\bibliography{Hippo_MaxEnt_Dec25}
\end{document}